# Efficient DNA-driven nanocavities for approaching quasi-deterministic strong coupling to a few fluorophores

Wan-Ping Chan[1], Jyun-Hong Chen[1], Wei-Lun Chou[1], Wen-Yuan Chen[1], Hao-Yu Liu[1], Hsiao-Ching Hu[3], Chien-Chung Jeng[2], Jie-Ren Li[3], Chi Chen[4], Shiuan-Yeh Chen[*1]

1. Department of Photonics, National Cheng Kung University, Tainan, Taiwan 70101
2. Department of Physics, National Chung Hsing University, Taichung, Taiwan 40227
3. Department of Chemistry, National Cheng Kung University, Tainan, Taiwan 70101
4. Research Center for Applied Science, Academia Sinica, Taipei, Taiwan 11529

*sychen72@ncku.edu.tw

**Abstract**

Strong coupling between light and matter is the foundation of promising quantum photonic devices such as deterministic single photon sources, single atom lasers and photonic quantum gates, which consist of an atom and a photonic cavity. Unlike atom-based systems, a strong coupling unit based on an emitter-plasmonic nanocavity system has the potential to bring these devices to the microchip scale at ambient conditions. However, efficiently and precisely positioning a single or a few emitters into a plasmonic nanocavity is challenging. In addition, placing a strong coupling unit on a designated substrate location is a demanding task. Here, fluorophore-modified DNA strands are utilized to drive the formation of particle-on-film plasmonic nanocavities and simultaneously integrate the fluorophores into the high field region of the nanocavities. High cavity yield and fluorophore coupling yield are demonstrated. This method is then combined with e-beam lithography to position the strong coupling units on designated locations of a substrate. Furthermore, the high correlation between electronic transition of the fluorophore and the cavity resonance is observed, implying more vibrational modes may be involved. Our system makes strong coupling units more practical on the microchip scale and at ambient conditions and provides a stable platform for investigating fluorophore-plasmonic nanocavity interaction.

Strong coupling between a single quantum emitter (QE) and a cavity provides a platform for quantum photonic devices[1-5]. Strong coupling between photonic cavities and QEs (i.e., atoms, molecules, quantum dots, J-aggregates) generate polariton modes (hybrid modes) and has been demonstrated in several systems which have evolved from atom-microwave cavity, to atom-optical cavity, to quantum dot-photonic crystal microcavity, and to QE-plasmonic nanocavity systems. In the past forty years, the size of the cavities involved in the strong coupling units substantially shrank by many orders of magnitude as the operating wavelength of the electromagnetic wave approached the optical regime[6,7].

While a single photonic cavity can simultaneously couple to many emitters, a strong coupling unit consisting of a single cavity and a single emitter is of particular interest due to their potential in single photon operations and nonlinear quantum photonic properties. The strong coupling units based on a single atom and a single cavity have been realized and applied to deterministic single photon sources, single atom lasers, and photonic quantum gates, etc.[1-4]

Although the atom-based strong coupling units have demonstrated proof-of-concept of novel quantum photonics devices, their bulky size due to atom manipulation methods (atom trapping, high vacuum) is inevitable. Realization of the strong coupling units on a solid state system such as a single quantum dot coupled to a photonic crystal cavity has made significant progress towards on-chip devices[5,7]. However, demanding precise alignment between a quantum dot and a photonic crystal microcavity is inevitable. In addition, due to the high-Q features of photonic crystal cavities for strong coupling, the corresponding quantum dots need to be controlled at cryogenic temperatures to fine-tune the resonant frequency.

In the past ten years, the plasmonic nanocavity with extremely small mode volumes has enabled cavity-QE strong coupling at ambient conditions. A plasmonic nanocavity coupled to many emitters[8-10], a few[11-13], and a single emitter[11-17] have been demonstrated. In addition, several theoretical works investigating QE-plasmonic cavity coupling have been conducted[18-24]. However, to efficiently and precisely position a single or a few emitters within a plasmonic nanocavity for constructing strong coupling units and to then place these units on the designated substrate locations are both challenging. Since the coupling strength is proportional to $1/\sqrt{V}$, the extremely small mode volume ($V$) associated with the plasmonic nanocavities facilitates the occurrence of strong coupling but also increases the difficulties of integrating an emitter into the cavity. Therefore, most plasmonic nanocavities coupled to single or a few emitters suffer from low cavity yield, low coupling yield and/or random distribution of the cavities[11,14,16]. In order to utilize the benefits of QE-plasmonic nanocavity systems, these drawbacks need to be eliminated.

In this report, fluorophore-modified DNA strands are utilized to construct a particle-on-film nanocavity and simultaneously integrate the emitters into the nanocavity using three distinct DNA configurations (1-Oligo, 2-Oligo and 3-Oligo). The 1-Oligo and 2-Oligo configurations are based on an affinity between DNA bases and gold film while the 3-Oligo configuration utilizes the hybridization of DNA strands from particle and film sides. In the results section, the high cavity yield and coupling yield are first shown. Second, one of the two efficient configurations (2-Oligo) is combined with e-beam lithography to position the strong coupling unit on the designated spots. The stability of the strong coupling units and the filling yield are shown. Third, the strong coupling is verified by photoluminescence (PL) and photobleaching tests. In the discussion section, the possibility of single fluorophore coupling in our system is discussed. The detailed coupling strength of the two efficient configurations are investigated. Finally, the detuning of the strong coupling units suggests that the transition frequency of the fluorophores may be highly correlated with the cavity's resonant frequency.

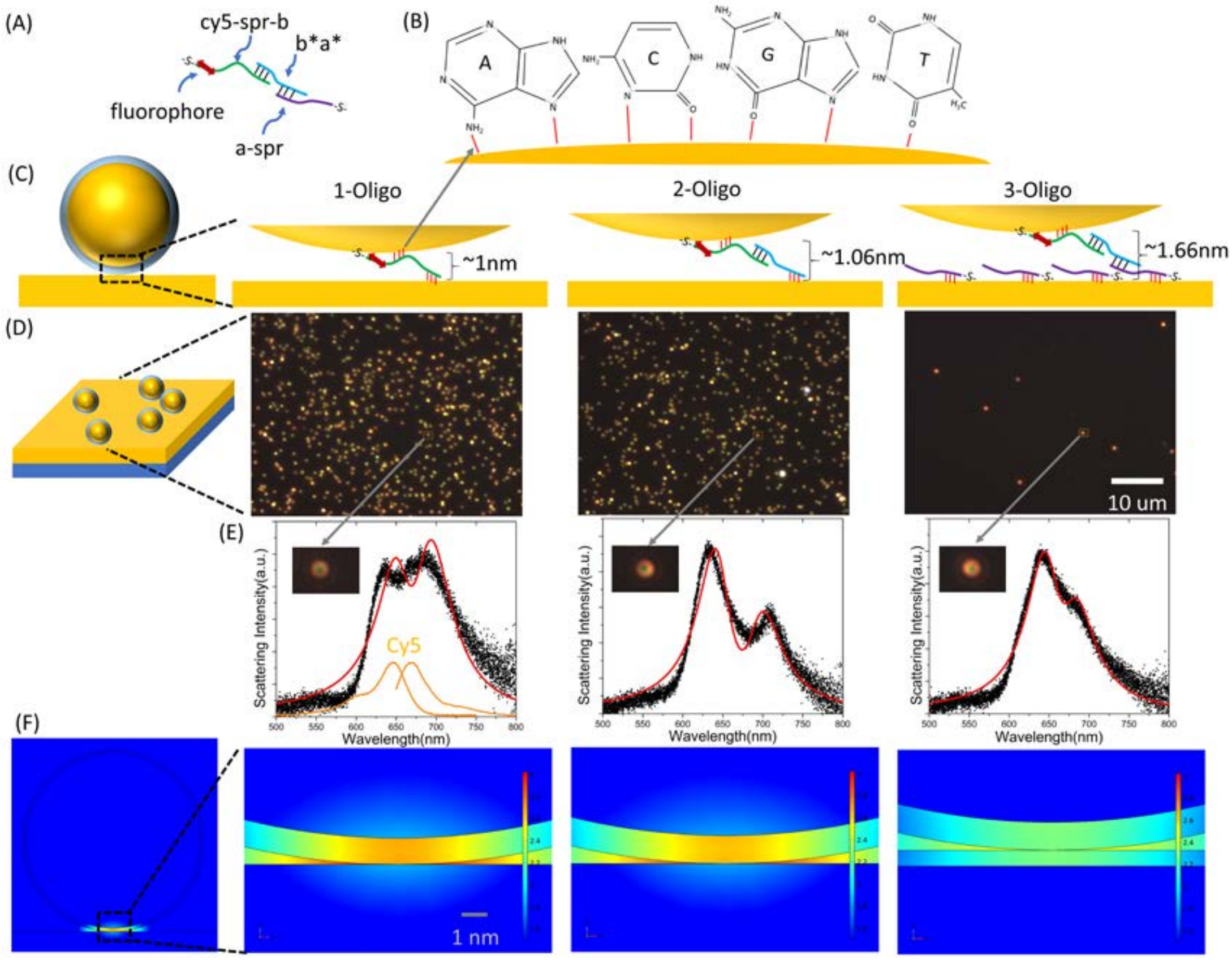

Fig. 1 (A) Illustration of the DNA strands and the fluorophore constructing particle-on-film nanocavities. (B) Affinity between DNA bases and gold surfaces is represented by the red lines. For the three DNA configurations: (C) Illustration of the three linkage configurations. (D) Dark-field images of the Cy5 embedded

nanocavities. (E) Scattering spectra (black dots) with theoretical fitting curves (red lines) and the corresponding images of the selected single nanocavities. The absorption and emission spectrum of Cy5 are also shown (orange curves). (F) Simulated near-field enhancement ($|E|/|E_{inc}|$) of the empty nanocavities (without fluorophores) on the log scale.

## Results

**The nanocavity formed by the affinity between nitrogen/oxygen atoms on the DNA bases and the gold substrate shows high cavity yield and high fluorophore coupling yield.** Three distinct DNA linkage configurations (1-Oligo, 2-Oligo and 3-Oligo) are utilized to simultaneously form the plasmonic nanocavities and integrate the fluorophores, Cy5, into the nanocavities. The relative position of the DNA strands (cy5-spr-b, b*a*, a-spr) used in these configurations is shown in Fig. 1(A). The detailed sequence of the three DNA strands are listed in supporting information 1. The "cy5" and "spr" represent the segments of the cy5 fluorophore and the spacer consisting of 10 adenine bases ($A_{10}$), respectively. The cy5-spr-b and a-spr strands are connected to the gold nanoparticles and gold films, respectively, through the thiol bond. The b* (a*) segment of b*a* strand is complementary to the b (a) segment of the cy5-spr-b (a-spr) strand. The affinity strength of the DNA bases for gold is A>C>G>T[25], Fig. 1(B). Therefore, the "spr" segment ($A_{10}$) can strongly adsorb onto the gold surface as a spacer separating the fluorophores. The three configurations (1-Oligo, 2-Oligo, and 3-Oligo) for constructing particle-on-film nanocavities are shown in Fig. 1(C). The strand cy5-spr-b is functionalized onto the gold nanoparticles in all three configurations. In the 1-Oligo configuration, a nanocavity is directly formed through the affinity between the strand cy5-spr-b and the gold film. Then, in the 2-Oligo and 3-Oligo configurations, the strand b*a* is added and partially hybridized with cy5-spr-b. A nanocavity is formed by the affinity between b*a* and the gold film in the 2-Oligo configuration. In the 3-Oligo configuration, a nanocavity is constructed by the hybridization between the strand b*a* and a-spr on the gold film.

The cavity yield is shown by the darkfield images (Fig. 1(D)) of particle-on-film nanocavities under three configurations. Each imaged dot corresponds to a single nanocavity. However, the whitish dots include several nanocavities with inter-distance below the diffraction limit. The nanocavities with the 1-Oligo and 2-Oligo configurations show at least one order of magnitude higher cavity yield (~200 cavities/(50 um)$^2$) than the 3-Oligo (~10 cavities/(50 um)$^2$). Please note that only cavities with red doughnut images are considered here because red doughnut images indicate that the desired gap dipole mode is dominant in the coupling to the fluorophores. The green/orange cavities result from plasmonic gap modes which are induced by imperfections on the nanosphere. These modes due to cavity imperfections can be clearly observed from the control (without Cy5) as well, and are discussed in several recent works[26-27]. In addition, the incubation time for the 1-Oligo, 2-Oligo configurations (30 s) is 120-fold shorter than the 3-Oligo configuration (1 hr). Therefore, contrast of the cavity yield would reach a difference of three orders of magnitude when the same incubation time is used. The low cavity yield of the 3-Oligo configuration may result from low hybridization efficiency due to the direct adsorption of the whole a-spr strand on the gold film. Adding supporting molecules for a-spr on the gold film may increase hybridization efficiency. In short, utilizing the affinity between DNA bases and gold can efficiently form the particle-on-film cavities. The high cavity yield facilitates the deposition of the strong coupling units on the designated spots of a substrate (Fig. 2).

Spectral splitting, the first signature of the coupling between the fluorophore exciton and the plasmonic nanocavity, is observed from ~30% of the nanocavities based on the 1-Oligo and 2-Oligo configurations (Fig. 1(E)). This coupling yield is high compared to related works (such as 1% in ref[16]). However, for the controls, which have the same DNA configurations except for Cy5, no splitting is observed (supporting information 2). This is the first evidence that the splitting results from fluorophore-plasmonic nanocavity coupling instead of other modes resulting from imperfections of the nanoparticles. On the contrary, no splitting is observed both in the 3-Oligo configuration and its control (supporting information 2). The reason will be discussed below.

For the 1-Oligo and 2-Oligo configurations, their similarity in cavity yield and ratio of spectral splitting reflects their comparable cavity properties, which are obviously different from the 3-Oligo configuration, as shown in Table 1. The average peak wavelength is obtained from the experimental spectra among the controls (without Cy5). The 1-Oligo and 2-Oligo configurations have resonance wavelengths closer to the emission peak of the Cy5. By assuming the refractive index of DNA is 1.455, which is a reasonable number comparable to experimental data from literature[28, 29], the estimated gap size is obtained by matching the simulated scattering peak to the experimental values. The 1-Oligo and 2-Oligo configurations have stronger field enhancements and smaller mode volumes than the 3-Oligo configuration. Since the coupling strength is proportional to $1/\sqrt{V}$, the Cy5-embedded nanocavities in the 1-Oligo and 2-Oligo configurations have a better indication of the strong coupling. Their fluorophore coupling yield and distribution of the coupling strength are investigated in the discussion section. The 1-Oligo and 2-Oligo configurations provide a similar cavity environment and comparable high coupling yield.

Table 1 The properties of the nanocavities formed by three configurations.

| | 1-Oligo | 2-Oligo | 3-Oligo |
|---|---|---|---|
| Empty nanocavity | | | |
| Average peak wavelength | 670 nm | 665 nm | 635 nm |
| Estimated gap size | 1 nm | 1.06 nm | 1.66 nm |
| Field enhancement at the center of the nanocavity | 470 | 429 | 180 |
| Mode volume ($nm^3$) | 80 | 94 | 311 |
| Cy5 embedded nanocavity | | | |
| Cavity yield ($\#/(50um)^2$) | 226 | 136 | 10 |
| Ratio of clear spectral splitting | 30% | 32% | NA |
| Fluorophore coupling yield | 92% | 82% | NA |

**The high cavity yield and coupling yield through the 1-Oligo and 2-Oligo configurations enable positioning the strong coupling units on the designated substrate locations.** In Fig. 2(A), the smooth gold film is covered with patterned PMMA holes fabricated by standard e-beam lithography. The PMMA wells with gold bottoms are incubated with the DNA-functionalized gold nanoparticles (2-Oligo configuration). After the particle-on-film cavity is formed, the pre-defined wells are removed. The Cy5 embedded nanocavities are left on the designated spot of the substrate.

Instead of randomly distributed cavities in Fig. 1(D), the cavities (Fig. 2(C)) showing clear mode splitting (Fig. 2(B)) are formed along the array pre-defined by 250 nm PMMA wells. The SEM images can confirm the actual number of cavities at each location (Fig. 2(D)). Because the diameter of the PMMA wells (250 nm) are 5 times larger than the nanoparticles (50 nm) in the sample shown in Fig. 2, some spots are occupied by more than a single cavity. The single-cavity spots usually have a red doughnut shape while the spots with multiple cavities are brighter and whiter (Fig. 2(C)). In addition, although the nanocavities are linked by the “soft” materials, DNA, after being stored in the ambient temperature and pressure conditions for one week, the scattering spectrum is still stable, Fig. 2(B). In Fig. 2(E), the filling yield is proportionally increased with the longer incubation time and larger diameter of the pattern wells. Here, the filling yield is defined by the number of cavity-occupied spots divided by the total number of spots regardless of whether the spot is occupied by a single particle or multiple particles. We observe that if the diameter of the pre-defined well is below 100 nm, only a single nanocavity is observed on each occupied spot. Please note that even at the same incubation condition, more time is required for depositing nanoparticles on the patterned gold than the bare gold surface in Fig. 1(D). The DNA-driven method is successfully utilized for positioning the strong coupling units on the designated substrate locations.

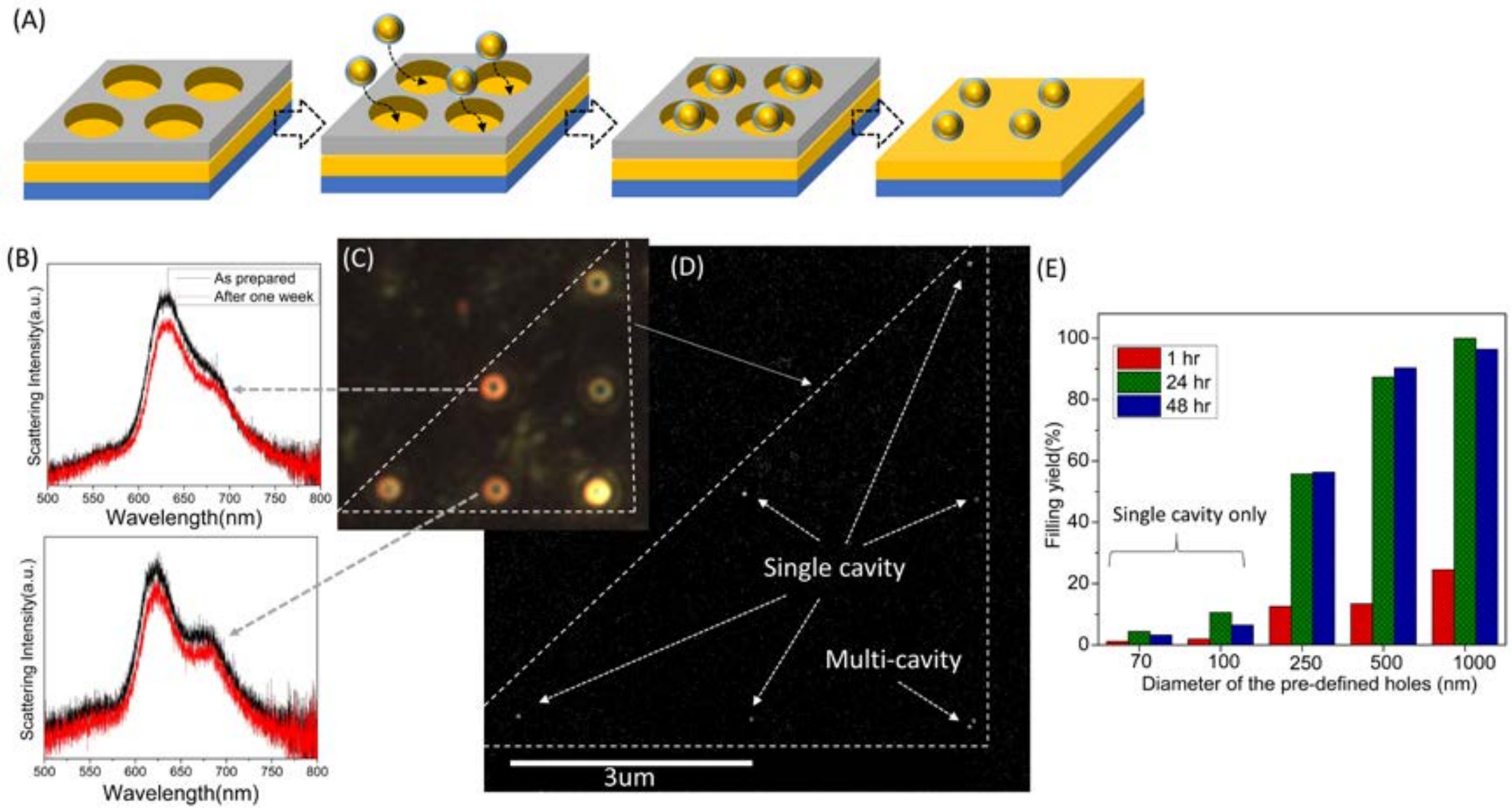

Fig. 2 Strong coupling units on the designated spots. (A) The fabrication process combining e-beam lithography and efficient affinity between DNA bases and the gold substrate. The scattering spectra (B), the dark-field image (C) and the corresponding SEM image (D) of the aligned cy5-embedded nanocavities. (E) The filling yield v.s. the size of pre-defined PMMA holes under three incubation times.

**Supporting evidence for the strong coupling.** The mode splitting observed in the scattering spectra from Cy5-embedded nanocavities rather than the controls clearly shows the coupling between Cy5 and nanocavities. However, recent reports suggest that the splitting in the scattering spectrum is insufficient for claiming the strong coupling between emitters and nanocavities[30-33]. In order to further confirm the coupling between the Cy5 exciton and the gap dipole mode of the particle-on-film nanocavity, photoluminescence (PL) spectrum[31-32,34-37] and photobleaching tests are both utilized[38].

The PL spectrum is used to probe the hybrid states of a single strong coupling unit since the pure absorption spectrum of a nanostructure is difficult to directly obtain. As shown in Fig. 3(A), the PL spectrum of a Cy5 embedded nanocavity (2-Oligo configuration) shows clear mode splitting as the scattering spectrum does. One the contrary, in Fig. 3(B), the single peak in the PL spectrum is observed from the control, which has been demonstrated by other groups and explained in terms of enhanced e−h pair generation and an increase in the emission rate both due to the gap-plasmon mode[39,40]. Furthermore, the photobleaching test is performed on the strong coupling units with continuous illumination using a 532 nm laser. Before the illumination, the entirety of the strong coupling unit is coated with polyelectrolyte layer by layer to increase the stability of the structure. Before and after each illumination, the scattering spectrum is measured. As shown in Fig. 3(C), the first 15 min of illumination do not cause the obvious change on the spectrum while the two scattering peaks are eliminated after the second illumination (60 min). The two peaks shown in the PL spectrum and eliminated in the scattering spectrum by the photobleaching test confirm the strong coupling between the Cy5 exciton and the nanocavity.

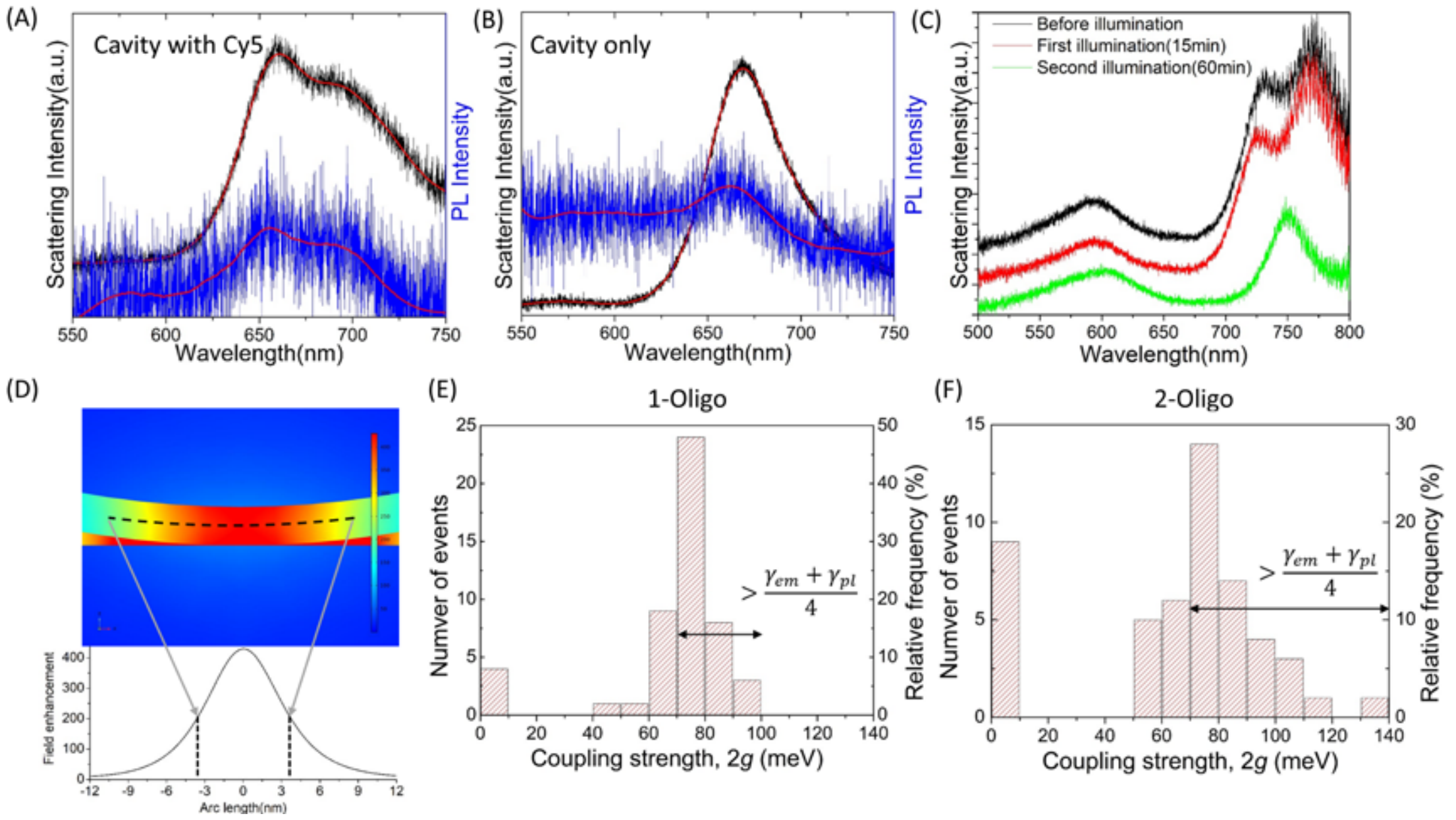

Fig. 3 Photoluminescence and scattering spectra of strong coupling units (A) and the control (B). (C) A photobleaching test monitored by the scattering spectra. (D) The electric field enhancement around the cavity region and line profile 0.5 nm away from the particle surface. The field enhancement plot is on a linear scale and saturated by the peak value of the line profile for clarity. The distribution of the coupling strength among Cy5-embedded nanocavities based on the 1-Oligo (E) and 2-Oligo (F) configurations.

## Discussion

**The possibility of strong coupling to a single Cy5 molecule.** Here, we use two approaches to discuss the possibility of strong coupling to a single Cy5 in our particle-on-film nanocavities. First, comparing the density of the molecules on the gold nanoparticle and spatial field distribution of the gold surface provides a rough estimation of the number of Cy5 in the hot spot region. Based on our previous report[41,42], the density of the Cy5 molecules on a 50 nm gold particle is 4 $nm^2$ per molecule. This density is obtained by measuring the unattached Cy5 in the supernatant during the salt aging process. This density is comparable to other literature[43]. As shown in Fig. 3(D), the amplitude of the electric field is concentrated in a 6 nm by 6 nm region. Therefore, the number of molecules within the cavity region is ~9, which makes the single Cy5 coupling possible.

Second, we can use eqn(1)[6] to estimate the number of molecules ($N$) involved in the coupling by the following quantities: coupling strength ($g$), dipole moments ($u_{em}$), and the amplitude of the vacuum field ($E_{vac}$). The coupling strength $g$=70 meV is obtained from the maxima splitting of the experimental spectrum fitted with the theoretical scattering cross-section ∝ $\omega^4 |u_{pl}|^2$ where $u_{pl}$ is the plasmon dipole, which is the solution of the coupled oscillator equations derived from semiclassical Maxwell-Bloch equations. The dipole moment of Cy5, $u_{em}$=15D is obtained from the transition frequency ($\nu_{em}$) and oscillation strength ($f$) which is derived from the integration of the absorption spectrum of Cy5, eqn(2)[44]. The $\varepsilon$ and $\bar{\nu}$ represents molar extinction coefficient and wave number, respectively. The $E_{vac}$ is expressed in terms of angular frequency of the nanocavity ($\omega_{pl}$), dielectric constant ($\epsilon_r$), permittivity of free space ($\varepsilon_0$), and mode volume ($V$), eqn(3)[45]. Based on the parameters described above, the number of molecules is estimated to be 0.5. This indicates that strong coupling to a single fluorophore in our system is possible. A recent theoretical work suggests that for particle-one-film cavities with a 1 nm gap, strong coupling can be achieved once the dipole moment of the emitter is ~10D[23]. Our results support their prediction.

$$g = \frac{\sqrt{N}\vec{\mu}_{em} \cdot \vec{E}_{vac}}{\hbar} \quad (1)$$

$$\vec{\mu}_{em} = \sqrt{\frac{3he^2}{8\pi m_e \nu_{em}} f} = \sqrt{\frac{3he^2}{8\pi m_e \nu_{em}}} \times (4.3 \times 10^{-9} \int \varepsilon d\bar{\nu}) \quad (2)$$

$$\vec{E}_{vac} = \sqrt{\frac{\hbar\omega_{pl}}{2\epsilon_r \varepsilon_0 V}} \quad (3)$$

**Quantitative analysis of the coupling between the fluorophore and nanocavity.** Here, we further investigate the coupling strength $g$ among the Cy5-embedded nanocavities and determine the ratio of them fulfilling the criterion for strong coupling, eqn(4)[47]. Since the criterion for strong coupling is not a hard limit, there are several criteria in literature[7,46,47]. Here, we choose the one in Pelton's recent review article which re-examines the important strong coupling works[47]. In some literature, this criterion is increased by a factor of 2. The $\gamma_{em}$=111 meV and $\gamma_{pl}$=145 meV represent the damping rate of the emitter and the cavity, respectively. Both values are extracted from the experimental spectra.

$$2g > \frac{\gamma_{em}+\gamma_{pl}}{4} \quad (4)$$

The coupling strength $g$ of the Cy5-embedded nanocavities is extracted from the fitting curves of the experimental scattering spectrum. The fitting curve is based on the coupled oscillator equations which are derived from the Maxwell-Bloch equation[48,49]. Fig. 3(E) and (F) show the distribution of $2g$ among 50 nanocavities in the 1-Oligo and 2-Oligo configurations, respectively. Both configurations have a similar $2g$ distribution: the mode is located at 70-80 meV, which is slightly larger than the strong coupling criterion (64 meV), and the small portion has no coupling strength. For the 1-Oligo configuration, 70% are above the strong coupling criterion, 22% are below the criterion, and 8% show no coupling. For the 2-Oligo configuration, 60% are above the strong coupling criterion, 22% are below the criterion, and 18% show no coupling. Therefore, the fluorophore coupling yield is at least >80% in the 1-Oligo and 2-Oligo configurations. It follows that if the orientation of the transition dipole can be further optimized, the strong coupling between Cy5 and the cavity would be more deterministic.

**The detuning of the strong coupling units.** In the conventional description of the light-matter strong coupling, the emitter has been modelled as a two-level system with a fixed transition energy $\hbar\omega_{em}$. Under this assumption, the hybrid polariton modes will follow a set of anti-crossing curves when the resonance of the cavity is detuned. However, we observe that when the Cy5-plasmonic nanocavity system (2-Oligo configuration) is detuned, the upper and lower hybrid modes follow a set of two parallel lines instead (blue and red dashed lines in Fig 4(A)). None of the three sets of anti-crossing curves based on the two-level model with distinct $\omega_{em}$ can fit the experimental data (blue and red solid lines in Fig. 4(A). The $\omega_{em}$, $\omega_{pl}$, and $g$ are extracted from the fitting curve of the experimental scattering spectrum (supporting information 3) and then used to calculate the upper $\omega_+$ and lower $\omega_-$ polariton modes by eqn(5)[6]. The $\omega_+$, $\omega_-$, and $\omega_{em}$ are represented by blue, red and black markers, respectively. Please note that the detuning of the same nanocavity uses the same markers. The regression line $\omega_{em} = 0.9143\omega_{pl} + 0.1549$ with correlation coefficient r=0.92 shows a highly positive correlation between $\omega_{em}$ and $\omega_{pl}$. If this linear regression relation is taken into eqn(5), two nearly parallel lines (blue and red dashed lines) will pass through the dataset consisting of $\omega_+$ and $\omega_-$.

$$\omega_{\pm} = \frac{1}{2}(\omega_{pl} + \omega_{em}) \pm \frac{1}{2}\sqrt{(2g)^2 + (\omega_{pl} - \omega_{em})^2} \quad (5)$$

We propose that the parallel polariton modes result from the mutable $\omega_{em}$ highly correlated with the frequency of the plasmonic nanocavity, $\omega_{pl}$. In Fig. 4(B)(C), the theoretical scattering curves are shown based on fixed $\omega_{em}$ and mutable $\omega_{em} = 0.9194\omega_{pl} + 0.1462$, respectively. The system with fixed $\omega_{em}$ shows the conventional anti-crossing curve while the mutable $\omega_{em}$ shows a set of parallel lines. Fieldman's group has experimentally and theoretically discovered that the fluorescence peak wavelength of the fluorophore is significantly correlated with the plasmonic resonance (linear regression line with the slope=0.95) in the weak coupling regime[49]. They suggest that an excited molecule tends to relax into a particular vibrational sublevel of the electronic ground state if the transition to that level is resonant with the plasmonic cavity. We suspect the same situation happens in the strong coupling regime. In addition, we suggest that more vibrational modes may be involved, which makes variation of $\omega_{em}$ more continuous than discretized. However, more theoretical studies are needed to clarify this experimental results in Fig. 4(A). Our system provides an efficient and stable experimental platform for the investigation of the fluorophore-nanocavity strong coupling.

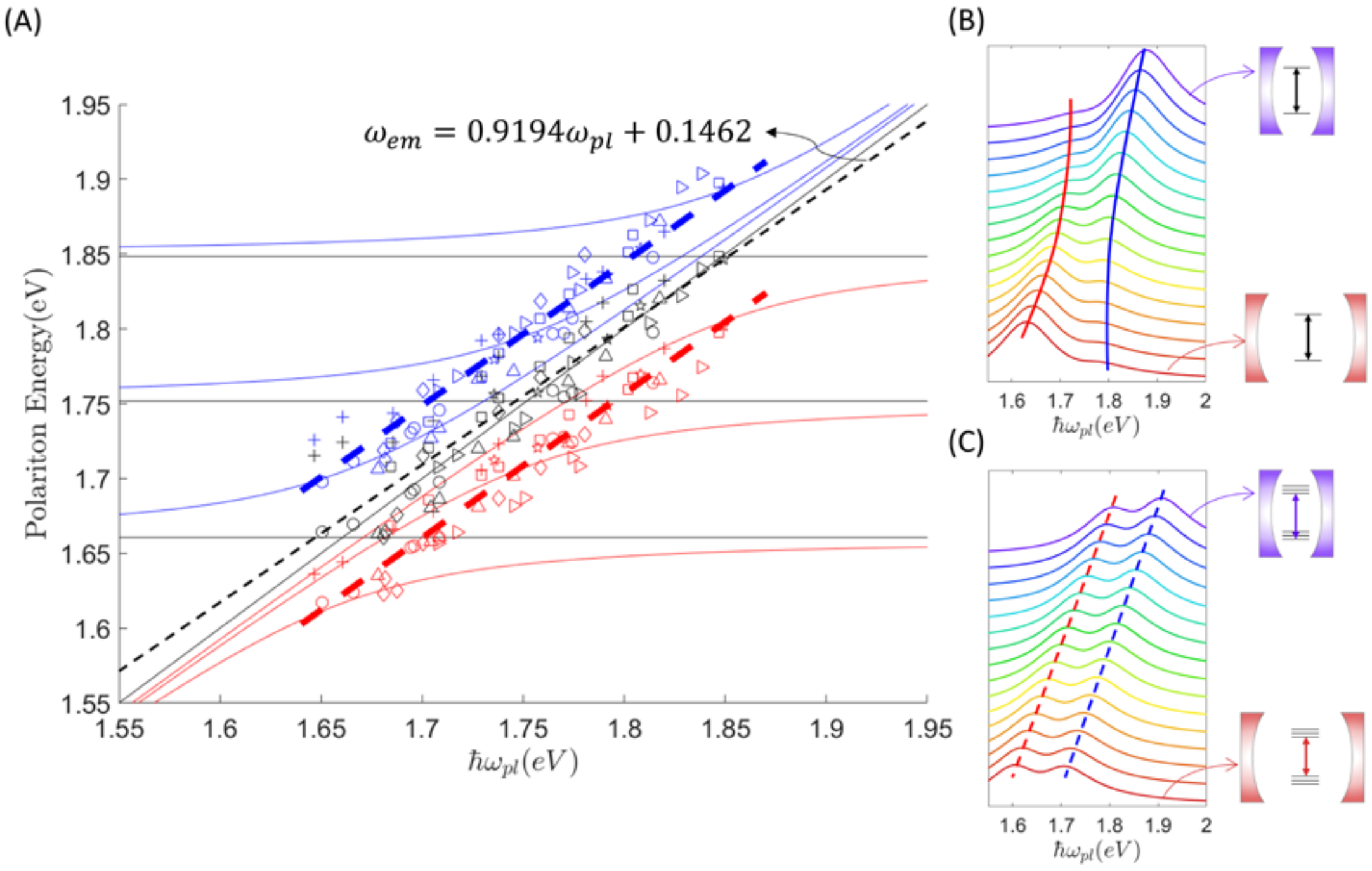

Fig. 4 (A) Detuning of the strong coupling units. Blue, red and black markers

represent the upper polariton, lower polariton modes and intrinsic electronic transitions of Cy5, respectively. The black dashed line is the regression line for ($\omega_{em}$, $\omega_{pl}$). The blue (red) dashed line is the upper (lower) polariton energy calculated with the regression relation of ($\omega_{em}$, $\omega_{pl}$) and eqn(5). Three sets of anti-crossing curves are separately derived from three distinct $\omega_{em}$ with eqn(5). (B) Calculated scattered spectra based on (B) fixed $\omega_{em}$ and (C) mutable $\omega_{em}$ under detuning of $\omega_{pl}$.

## Conclusion

In this work, the strong coupling units consisting of fluorophores and plasmonic nanocavities are constructed by an efficient DNA-driven method, resulting in high cavity yield and high coupling yield. This method is further applied to the patterned gold substrate so that the strong coupling units can be positioned on the designated locations. The splitting in scattering/PL spectrum and elimination of the splitting in the photobleaching tests clearly demonstrate the coupling between the Cy5 exciton and the nanocavity. Comparing the spatial distribution of the field enhancement and the density of Cy5 suggests that the number of Cy5 within the cavity is ~9. Based on the experimental coupling strength, the transition dipole moment, and the calculated mode volume, the single Cy5 coupling in our system is possible. The detuning of Cy5-embedded nanocavities shows a set of parallel curves, which probably results from highly correlated $\omega_{em}$ and $\omega_{pl}$. Their high correlation facilitates the design for fluorophore-based quantum photonics devices. The coupling strength and coupling yield may be further improved by a different DNA design. Our results provide an efficient way to construct a molecule-based strong coupling unit, which may be applied to on-chip quantum photonic devices at ambient temperature and pressure. It also provides a platform for studying molecule-nanocavity interaction.

## Method

**DNA modified gold nanoparticles/films for nanocavity formation.** First, for all three configurations, the 50 nm gold nanoparticles (BBI Inc.) are coated with Cy5 modified DNA strands (IDT-DNA), cy5-spr-b, through the salt aging method. The detailed steps are described in our previous report[41,42]. Then, for the 2-Oligo and 3-Oligo configurations, the strand b*a* is added to the solution of Cy5-spr-b modified Au nanoparticles for b*b hybridization. The temperature of the mixed solution is increased to 60℃, maintained for one hour, and then annealed to the ambient temperature. For the 3-Oligo configuration, the gold film is also coated with strand spr-a through immersing the film in a solution of spr-a stands for 24 hr. For the 1-Oligo and 2-Oligo configurations, the freshly-peeled gold films are incubated in solutions of gold nanoparticles coated with the respective DNA strands for 30 s. For the 3-Oligo configuration, the spr-a functionalized gold film is immersed in a solution of gold nanoparticles with DNA coating for 1 hr due to lower hybridization efficiency.

**Smooth gold film preparation.** The 90 nm gold film is deposited on freshly cleaved mica by thermal evaporation at a deposition rate of 1 A/s. Then, the quartz substrates are attached to the gold film by glue. The fresh gold surface is obtained by peeling the gold off the mica. The surface morphology of the gold film is shown in supporting information 4.

**Nanocavities on the designed spots.** The patterned gold surface is fabricated by adding the PMMA wells on smooth gold film. The PMMA wells are fabricated by e-beam lithography. Five different diameters (70 nm, 100 nm, 250 nm, 500 nm, 1 um ) of wells with 70 nm depth are fabricated. The patterned gold film is immersed in a solution of gold nanoparticles with a DNA coating (2-Oligo) for three incubation times (1 hr, 24 hr, 48 hr). The gold nanoparticles are deposited on the gold-bottomed surface of the PMMA wells. After the nanocavities are formed on the designated spots, the PMMA wells are removed by immersing the sample in a mixed solution of isopropanol:water (4:1) for 4 hr.

**Scattering/photoluminescence spectrum and photobleaching.** The scattering spectrum of a Cy5-embedded nanocavity is collected by a custom microscope under a dark-field imaging configuration. The nanocavity is illuminated by a 100 W halogen lamp through a 100x objective, and the scattered light is collected by the same objective and relayed to the spectrometer. The size of the slit is open to 300 um and the grating is 600 g/mm. For the photoluminescence spectrum, the excitation source is changed to a 532 nm cw laser at 100 W/cm$^2$ intensity at the output of the objective. The PL signals are filtered out from the strong elastic scattering by a long pass filter. The photobleaching tests for Cy5 are performed by illuminating the Cy5 embedded nanocavities with a 532 nm cw laser (100 W/cm$^2$) for 15 min (first illumination) and 60 min (second illumination).

**Polyelectrolyte coating on the nanocavities.** In order to finely tune the resonant wavelength of the nanocavity for the detuning experiments or enhance the structural stability during the photobleaching tests, two polyelectrolytes with opposing polarity are used to coat the nanocavities by alternately immersing the sample into solutions of 1 M NaCl + 3 mM PAH (polyallylamine hydrochloride) and 1M NaCl+ 3 mM PSS (polystyrene sulfonate) for 15 min for each coating[50].

**Electromagnetic simulation.** The near-field distribution (field enhancement) of the nanoparticle-on-film nanocavity is calculated by the finite element method in the frequency domain (COMSOL 5.5, RF module). The p-pol excitation at 70° to the normal is used. The mode volume is calculated based on the following equation[11]:

$$V(\omega) = \frac{\int W(r,\omega) d^3 r}{\max\,[W(r,\omega)]}$$

where

$$W(r,\omega) = \frac{1}{2}\left(\frac{\partial[\omega\varepsilon(r,\omega)]}{\partial\omega}\varepsilon_0|E(r,\omega)|^2 + \mu_0|H(r,\omega)|^2\right)$$

## Acknowledgements

S.-Y Chen would like to thank Dr. V. Mao for manuscript review, J.-Y. Chen for programming assistance, and Dr. L.-Y. Hsu, Dr. B.-L. Huang, Prof. Y.-C. Chang, Prof. C-S. Chuu for useful discussions. This project is supported by the Ministry of Science and Technology, Taiwan (MOST 109-2221-E-006-205, MOST 108-2221-E-006-205, MOST 107-2221-E-006-147, MOST 106-2221-E-006-171)

## Author contributions

S.-Y. C. designed the research and wrote the manuscript. C. C., J.-R. L., H.-C. H., W.-L. C. contributed gold film preparation and characterization. C.-C. J. contributed the optical setup. H.-Y. L. contributed electromagnetic simulations. W.-P. C. J.-H. C., W.-L. C., W.-Y. C. performed the experiments. W.-P. C. and J.-H. C. contributed the data analysis. S.-Y. C., W.-P. C. and C. C. revised the manuscript.

## References

1. McKeever, J., Boca, A., Boozer, A. D., Buck, J. R. &Kimble, H. J. Experimental realization of a one-atom laser in the regime of strong coupling. *Nature* **425**, 268–271 (2003).
2. McKeever, J. *et al.* Deterministic Generation of Single Photons from One Atom Trapped in a Cavity. *Science* **303**, 1992–1994 (2004).
3. Reiserer, A., Kalb, N., Rempe, G. &Ritter, S. A quantum gate between a flying optical photon and a single trapped atom. *Nature* **508**, 237–240 (2014).
4. Hacker, B., Welte, S., Rempe, G. &Ritter, S. A photon–photon quantum gate based on a single atom in an optical resonator. *Nature* **536**, 193–196 (2016).
5. He, Y.-M. *et al.* On-demand semiconductor single-photon source with near-unity indistinguishability. *Nat. Nanotechnol.* **8**, 213–217 (2013).
6. Törmä, P. &Barnes, W. L. Strong coupling between surface plasmon polaritons and emitters: a review. *Reports Prog. Phys.* **78**, 13901 (2014).
7. Khitrova, G., Gibbs, H. M., Kira, M., Koch, S. W. &Scherer, A. Vacuum Rabi splitting in semiconductors. *Nat. Phys.* **2**, 81–90 (2006).
8. Schlather, A. E., Large, N., Urban, A. S., Nordlander, P. &Halas, N. J. Near-field mediated plexcitonic coupling and giant Rabi splitting in individual metallic dimers. *Nano Lett.* **13**, 3281–3286 (2013).
9. Zengin, G. *et al.* Approaching the strong coupling limit in single plasmonic nanorods interacting with J-aggregates. *Sci. Rep.* **3**, 1–8 (2013).
10. Zengin, G. *et al.* Realizing strong light-matter interactions between single-nanoparticle plasmons and molecular excitons at ambient conditions. *Phys. Rev. Lett.* **114**, 157401 (2015).
11. Chikkaraddy, R. *et al.* Single-molecule strong coupling at room temperature in plasmonic nanocavities. *Nature* **535**, 127–130 (2016).
12. Liu, R. *et al.* Strong light-matter interactions in single open plasmonic nanocavities at the quantum optics limit. *Phys. Rev. Lett.* **118**, 237401 (2017).
13. Luo, Y. *et al.* Deterministic coupling of site-controlled quantum emitters in monolayer WSe 2 to plasmonic nanocavities. *Nat. Nanotechnol.* **13**, 1137–1142 (2018).
14. Santhosh, K., Bitton, O., Chuntonov, L. &Haran, G. Vacuum Rabi splitting in a plasmonic cavity at the single quantum emitter limit. *Nat. Commun.* **7**, 1–5 (2016).
15. Groß, H., Hamm, J. M., Tufarelli, T., Hess, O. &Hecht, B. Near-field strong coupling of single quantum dots. *Sci. Adv.* **4**, eaar4906 (2018).
16. Leng, H., Szychowski, B., Daniel, M.-C. &Pelton, M. Strong coupling and induced transparency at room temperature with single quantum dots and gap plasmons. *Nat. Commun.* **9**, 1–7 (2018).

17. Ojambati, O. S. *et al.* Quantum electrodynamics at room temperature coupling a single vibrating molecule with a plasmonic nanocavity. *Nat. Commun.* **10**, 1–7 (2019).
18. Salomon, A., Gordon, R. J., Prior, Y., Seideman, T. &Sukharev, M. Strong coupling between molecular excited states and surface plasmon modes of a slit array in a thin metal film. *Phys. Rev. Lett.* **109**, 73002 (2012).
19. Yang, J., Perrin, M. &Lalanne, P. Analytical formalism for the interaction of two-level quantum systems with metal nanoresonators. *Phys. Rev. X* **5**, 21008 (2015).
20. Galego, J., Garcia-Vidal, F. J. &Feist, J. Cavity-induced modifications of molecular structure in the strong-coupling regime. *Phys. Rev. X* **5**, 41022 (2015).
21. Li, R.-Q., Hernángomez-Pérez, D., García-Vidal, F. J. &Fernández-Domínguez, A. I. Transformation optics approach to plasmon-exciton strong coupling in nanocavities. *Phys. Rev. Lett.* **117**, 107401 (2016).
22. Kewes, G., Binkowski, F., Burger, S., Zschiedrich, L. &Benson, O. Heuristic modeling of strong coupling in plasmonic resonators. *ACS Photonics* **5**, 4089–4097 (2018).
23. Rousseaux, B., Baranov, D. G., Käll, M., Shegai, T. &Johansson, G. Quantum description and emergence of nonlinearities in strongly coupled single-emitter nanoantenna systems. *Phys. Rev. B* **98**, 45435 (2018).
24. Zhao, D. *et al.* Impact of vibrational modes in the plasmonic Purcell effect of organic molecules. *ACS photonics* **7**, 3369–3375 (2020).
25. Koo, K. M., Sina, A. A. I., Carrascosa, L. G., Shiddiky, M. J. A. &Trau, M. DNA–bare gold affinity interactions: mechanism and applications in biosensing. *Anal. Methods* **7**, 7042–7054 (2015).
26. Huh, J.-H., Lee, J. &Lee, S. Comparative study of plasmonic resonances between the roundest and randomly faceted Au nanoparticles-on-mirror cavities. *ACS Photonics* **5**, 413–421 (2018).
27. Kongsuwan, N. *et al.* Plasmonic nanocavity modes: From near-field to far-field radiation. *ACS Photonics* **7**, 463–471 (2020).
28. Elhadj, S., Singh, G. &Saraf, R. F. Optical properties of an immobilized DNA monolayer from 255 to 700 nm. *Langmuir* **20**, 5539–5543 (2004).
29. Liu, G. L. *et al.* A nanoplasmonic molecular ruler for measuring nuclease activity and DNA footprinting. *Nat. Nanotechnol.* **1**, 47–52 (2006).
30. Antosiewicz, T. J., Apell, S. P. &Shegai, T. Plasmon–exciton interactions in a core–shell geometry: from enhanced absorption to strong coupling. *ACS*

*Photonics* **1**, 454–463 (2014).

31. Zengin, G. *et al.* Evaluating conditions for strong coupling between nanoparticle plasmons and organic dyes using scattering and absorption spectroscopy. *J. Phys. Chem. C* **120**, 20588–20596 (2016).
32. Stete, F., Koopman, W. &Bargheer, M. Signatures of strong coupling on nanoparticles: revealing absorption anticrossing by tuning the dielectric environment. *ACS Photonics* **4**, 1669–1676 (2017).
33. Kongsuwan, N. *et al.* Quantum plasmonic immunoassay sensing. *Nano Lett.* **19**, 5853–5861 (2019).
34. Wersäll, M. *et al.* Correlative dark-field and photoluminescence spectroscopy of individual plasmon–molecule hybrid nanostructures in a strong coupling regime. *ACS photonics* **6**, 2570–2576 (2019).
35. Wersall, M., Cuadra, J., Antosiewicz, T. J., Balci, S. &Shegai, T. Observation of mode splitting in photoluminescence of individual plasmonic nanoparticles strongly coupled to molecular excitons. *Nano Lett.* **17**, 551–558 (2017).
36. Zhou, N., Yuan, M., Gao, Y., Li, D. &Yang, D. Silver nanoshell plasmonically controlled emission of semiconductor quantum dots in the strong coupling regime. *ACS Nano* **10**, 4154–4163 (2016).
37. Melnikau, D. *et al.* Rabi splitting in photoluminescence spectra of hybrid systems of gold nanorods and J-aggregates. *J. Phys. Chem. Lett.* **7**, 354–362 (2016).
38. Stete, F., Schoßau, P., Bargheer, M. &Koopman, W. Size-Dependent Coupling of Hybrid Core–Shell Nanorods: Toward Single-Emitter Strong-Coupling. *J. Phys. Chem. C* **122**, 17976–17982 (2018).
39. Lumdee, C., Yun, B. &Kik, P. G. Gap-plasmon enhanced gold nanoparticle photoluminescence. *ACS Photonics* **1**, 1224–1230 (2014).
40. Olesiak-Banska, J., Waszkielewicz, M., Obstarczyk, P. &Samoc, M. Two-photon absorption and photoluminescence of colloidal gold nanoparticles and nanoclusters. *Chem. Soc. Rev.* **48**, 4087–4117 (2019).
41. Chen, S.-Y. &Lazarides, A. A. Quantitative amplification of Cy5 SERS in 'warm spots' created by plasmonic coupling in nanoparticle assemblies of controlled structure. *J. Phys. Chem. C* **113**, 12167–12175 (2009).
42. Chen, S. Y. *et al.* Gold nanoparticles on polarizable surfaces as raman scattering antennas. *ACS Nano* **4**, 6535–6546 (2010).
43. Hill, H. D., Millstone, J. E., Banholzer, M. J. &Mirkin, C. A. The role radius of curvature plays in thiolated oligonucleotide loading on gold nanoparticles. *ACS Nano* **3**, 418–424 (2009).
44. Haken, H. &Wolf, H. C. *Molecular physics and elements of quantum chemistry*.

(Springer, 1995).

45. Yamamoto, Y., Tassone, T. &Cao, H. *Semiconductor cavity quantum electrodynamics.* (Springer, 2000).

46. Chikkaraddy, R. *et al.* Mapping nanoscale hotspots with single-molecule emitters assembled into plasmonic nanocavities using DNA origami. *Nano Lett.* **18**, 405–411 (2018).

47. Pelton, M., Storm, S. D. &Leng, H. Strong coupling of emitters to single plasmonic nanoparticles: exciton-induced transparency and Rabi splitting. *Nanoscale* **11**, 14540–14552 (2019).

48. Shah, R. A., Scherer, N. F., Pelton, M. &Gray, S. K. Ultrafast reversal of a Fano resonance in a plasmon-exciton system. *Phys. Rev. B* **88**, 75411 (2013).

49. Ringler, M. *et al.* Shaping emission spectra of fluorescent molecules with single plasmonic nanoresonators. *Phys. Rev. Lett.* **100**, 203002 (2008).

50. Huang, C.-Z., Wu, M.-J. &Chen, S.-Y. High order gap modes of film-coupled nanospheres. *J. Phys. Chem. C* **119**, (2015).

**Efficient DNA-driven nanocavities for approaching quasi-deterministic strong coupling to a few fluorophores**

Wan-Ping Chan[1], Jyun-Hong Chen[1], Wei-Lun Chou[1], Wen-Yuan Chen[1], Hao-Yu Liu[1], Hsiao-Ching Hu[3], Chien-Chung Jeng[2], Jie-Ren Li[3], Chi Chen[4], Shiuan-Yeh Chen[*1]
1. Department of Photonics, National Cheng Kung University, Tainan, Taiwan 70101
2. Department of Physics, National Chung Hsing University, Taichung, Taiwan 40227
3. Department of Chemistry, National Cheng Kung University, Tainan, Taiwan 70101
4. Research Center for Applied Science, Academia Sinica, Taipei, Taiwan 11529
*sychen72@ncku.edu.tw

# Supporting Information

S1. The sequence of the DNA strands used in this work. 1-Oligo configuration uses cy5-spr-b only. 2-Oligo configuration uses cy5-spr-b and b*a*. 3-Oligo configuration uses cy5-spr-b, b*a*, and a-spr. The strand spr-b replaces cy5-spr-b for the corresponding controls.

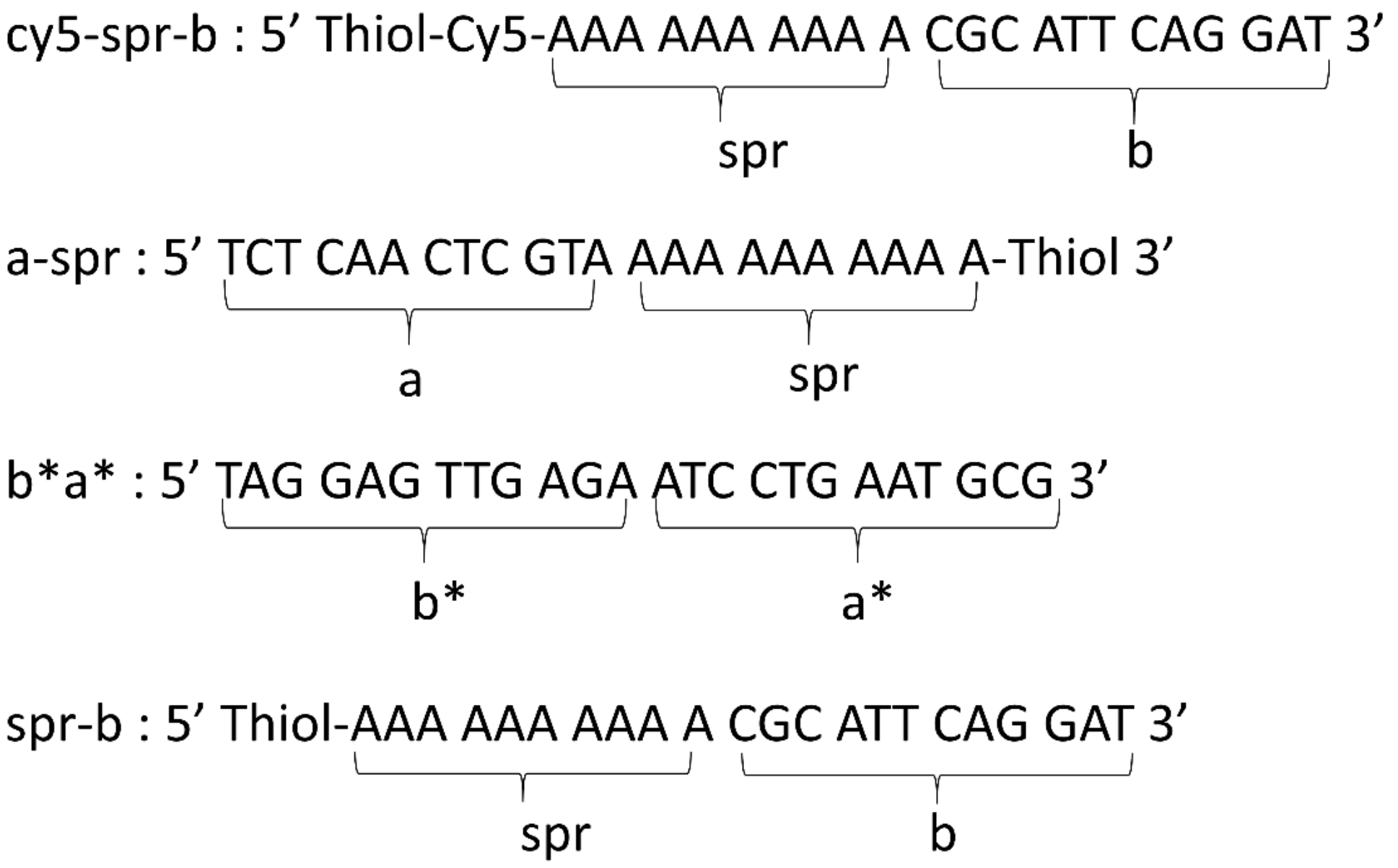

Fig. S1 The sequence of the DNA strands.

S2. The scattering images and spectra of the controls.

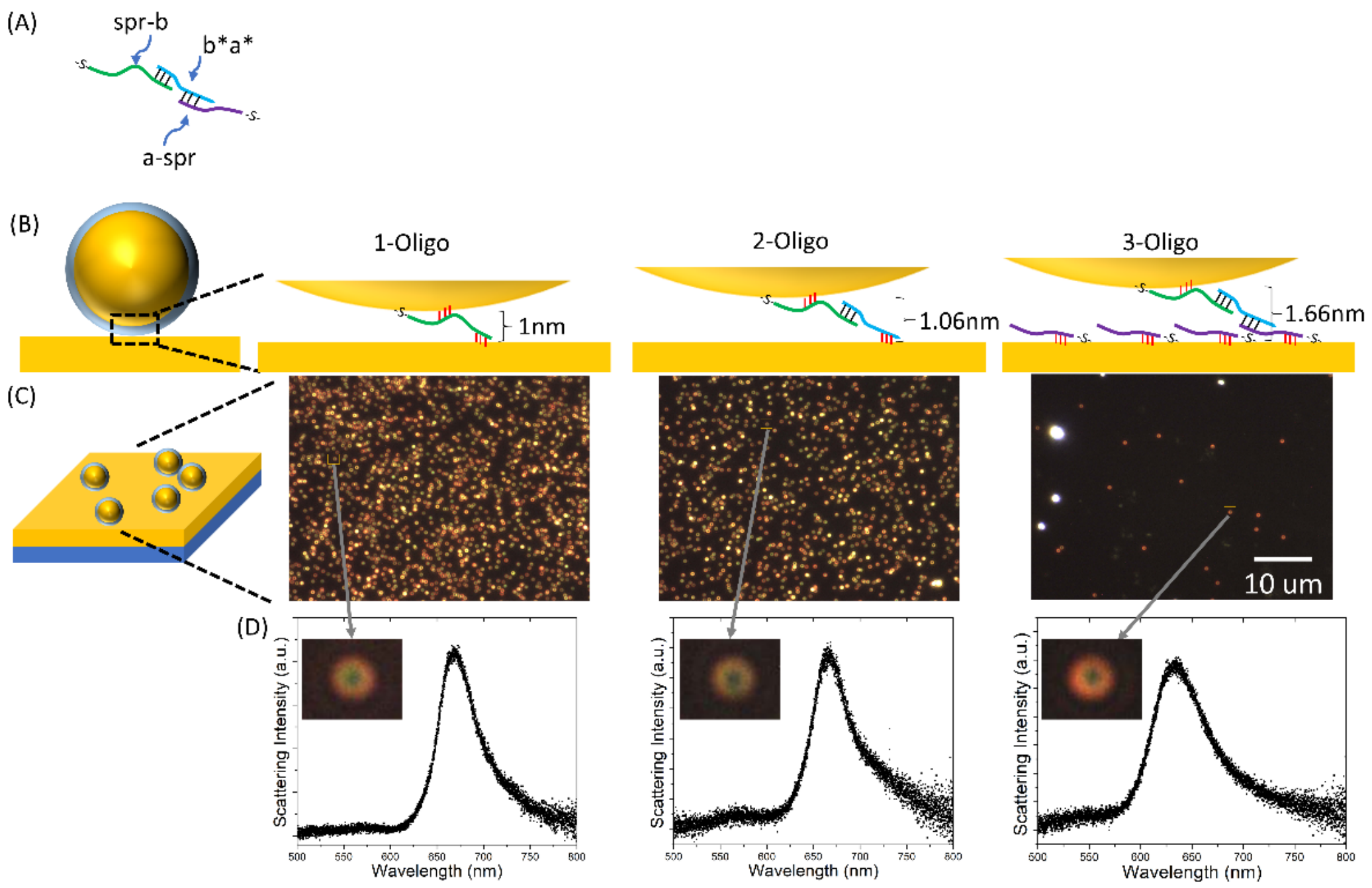

Fig. S2 (A) The illustration of DNA strands used for the controls. For three DNA configurations: (B) Illustration of the three linkage configurations. (C) Dark-field images of the nanocavities. (D) Scattering spectra and the corresponding images of the selected single nanocavities.

S3. The experimental scattering spectra of the Cy5-embedded nanocavities under detuning in Fig. 4(A).

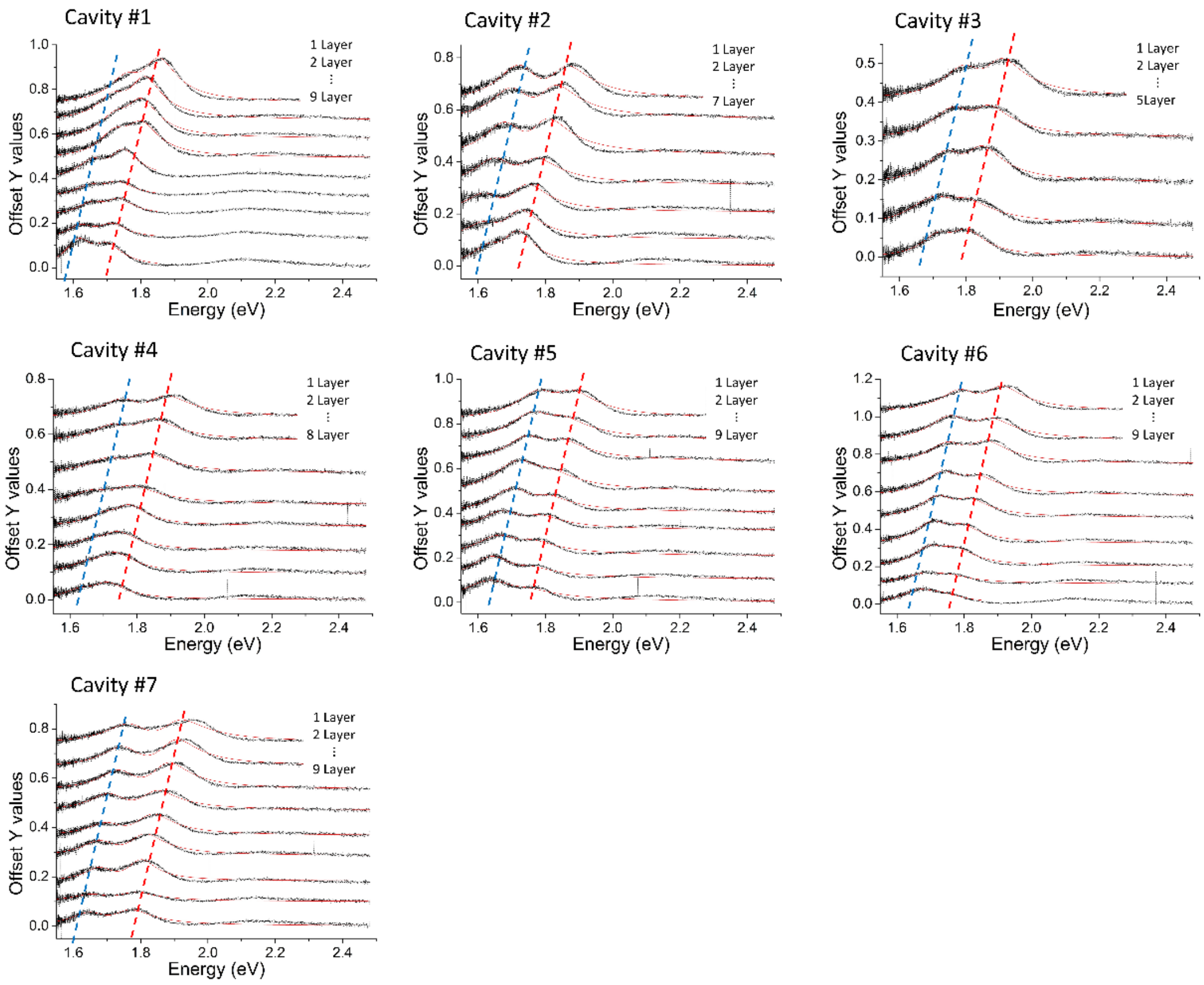

Fig. S3 The scattering spectra of seven Cy5-embedded nanocavities. The red curves are the corresponding fitting curves. The blue and red dash lines are for eye-guidance.

S4. The surface morphology of the smooth gold film.

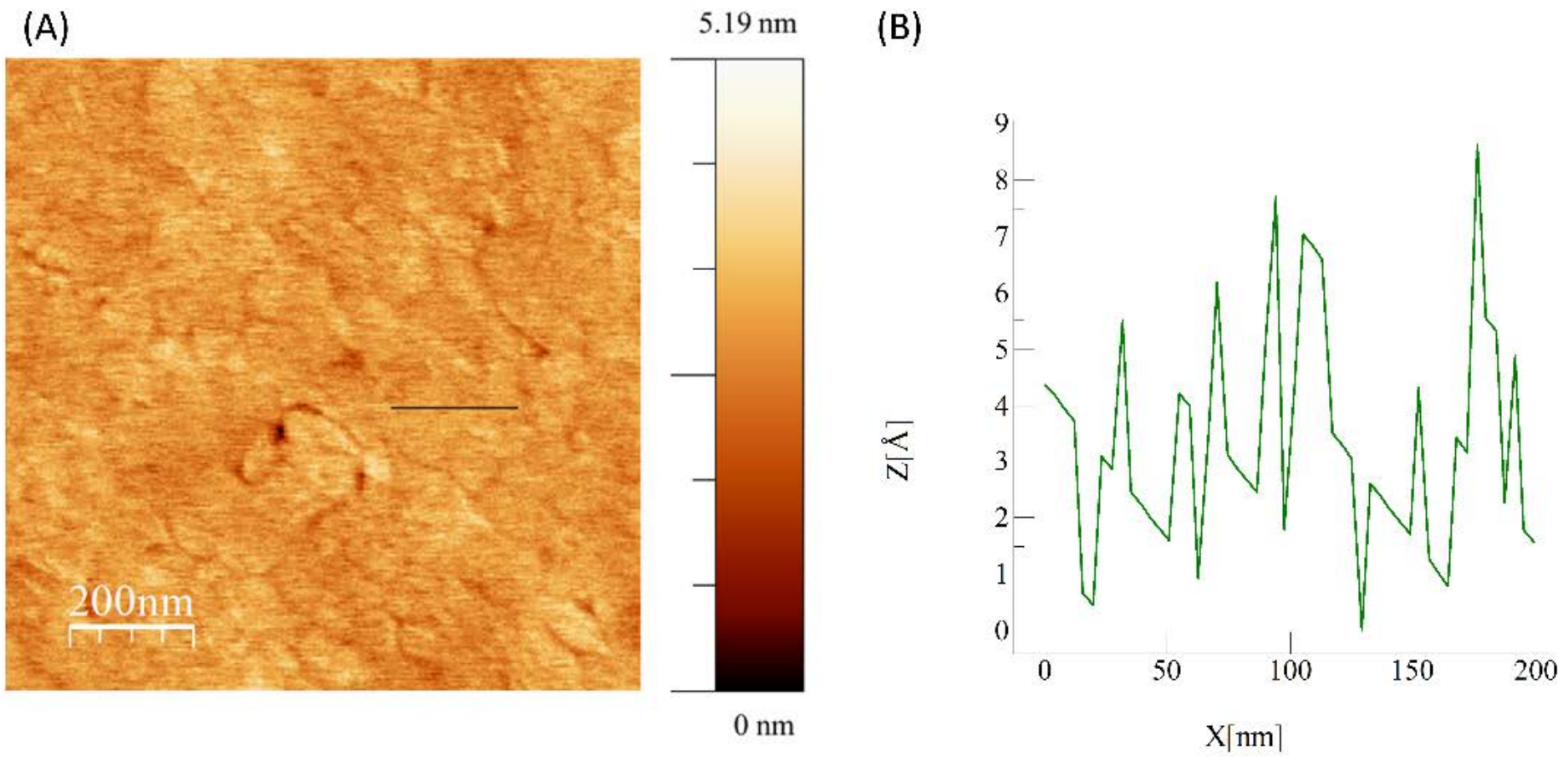

Fig. S4 The AFM image (A) and the height profile (B) of the smooth gold film.